\documentclass[12pt]{article}
\usepackage{graphicx}

\begin{document}

\title{A thermal coherent state defined with the Lie-Trotter product formula}

\author{Hiroo Azuma${}^{1,}$\thanks{Email: hiroo.azuma@m3.dion.ne.jp}
\ \ 
and
\ \ 
Masashi Ban${}^{2,}$\thanks{Email: m.ban@phys.ocha.ac.jp}
\\
\\
{\small ${}^{1}$Advanced Algorithm \& Systems Co., Ltd.,}\\
{\small 7F Ebisu-IS Building, 1-13-6 Ebisu, Shibuya-ku, Tokyo 150-0013, Japan}\\
{\small ${}^{2}$Graduate School of Humanities and Sciences, Ochanomizu University,}\\
{\small 2-1-1 Ohtsuka, Bunkyo-ku, Tokyo 112-8610, Japan}
}

\date{\today}

\maketitle

\begin{abstract}
In this paper,
we investigate a thermal coherent state defined with the Lie-Trotter product formula
under the formalism of the thermo field dynamics.
In the definition of our thermal coherent state,
we treat the thermalizing operator and the displacement operator symmetrically.
We examine its uncertainty relation and quasiprobability distributions.
Although this thermal coherent state is equivalent to the conventional ones except for different parameterizations and a phase factor,
it is convenient for describing an experimental setup
of
the optical parametric oscillator laser.
\end{abstract}

\section{\label{section-introduction}Introduction}
In order to extend the coherent state to finite temperature,
Barnett, Knight, Garavaglia, Mann and Revzen introduce the thermal coherent states
using the formalism of the thermo field dynamics (TFD)
\cite{Barnett1985,Garavaglia1988,Mann1989a}.
They define their thermal coherent states by applying the thermalizing operator and the displacement operator
for free bosons
to the zero-temperature vacuum state.
Thus, so far,
there have been two natural candidates for the thermal coherent states
according to the order of the thermalizing operator and the displacement operator.

In Refs.~\cite{Mann1989b,Ezawa1991,Oz-Vogt1991},
the characteristic functions of the thermal coherent states are derived.
Their uncertainty relations are discussed in Ref.~\cite{Mann1989b}.
Time evolution of the Jaynes-Cummings model
with the thermal coherent state at initial time is examined in Ref.~\cite{Azuma2011}.
We can expect that the thermal coherent states are convenient for analyzing quantum mechanical systems
at finite temperature.

In this paper,
we propose a thermal coherent state defined with the Lie-Trotter product formula.
We investigate its properties from both theoretical and experimental viewpoints.
To define our thermal coherent state,
we treat the thermalizing operator and the displacement operator symmetrically.
In the former half of this paper,
we show that this thermal coherent state is equivalent to the conventional ones except for different parameterizations
and a phase factor.
Moreover,
we examine its uncertainty relation and quasiprobability distributions.
In the latter half of this paper,
we consider how a physical system actually realizes our thermal coherent state.
We examine an experimental setup
of
the optical parametric oscillator (OPO) laser system.
We explain how this physical system induces our thermal coherent state in the laboratory.
Thus, we can conclude that our thermal coherent state is convenient for describing quantum states of photons under a real experiment.
This is the motivation for the current paper.

This paper is organized as follows.
In the latter half of this section,
we give brief reviews of
the formalism of the TFD and the Lie-Trotter product formula,
both of which play important roles in this paper.
In Sec.~\ref{section-definition-thermal-coherent-states},
we give a definition of a new thermal coherent state
using the Lie-Trotter product formula.
We prove that it is equivalent to conventional thermal coherent states except for different parameterizations and a phase factor.
In Sec.~\ref{section-properties-thermal-coherent-state},
we derive an uncertainty relation and quasiprobability distributions of our thermal coherent state.
In Sec.~\ref{section-OPO-laser},
we examine an experimental setup
of the OPO laser
that actually realizes our thermal coherent state
in the laboratory.
In Sec.~\ref{section-discussions},
we give brief discussions.
In Appendix~\ref{section-Appendix-A},
we show that two conventional thermal coherent states are equivalent to each other.
Moreover,
we derive their uncertainty relations and characteristic functions.
In Appendix~\ref{section-Appendix-B},
we evaluate quasiprobability distributions of conventional thermal coherent states.

Here,
first, we explain the TFD \cite{Takahashi1975,Umezawa1982,Umezawa1992}.
The TFD is a formulation for describing quantum mechanical systems at finite temperature.
Using this formalism,
we can calculate a statistical average of an observable at finite temperature
as an expectation value of a pure state.
Thus,
under the TFD,
we do not need to think about mixed state,
and this is an advantage of the TFD.

The price of this convenience is that we have to introduce a tilde particle
corresponding to an ordinary particle in the TFD.
Then,
the ordinary particles and the tilde particles represent the dynamical degree of freedom
and the thermal degree of freedom,
respectively.
Thus,
to every state $|n\rangle$ belonging to the original Hilbert space ${\cal H}$,
we associate $|\tilde{n}\rangle$ belonging to the tilde conjugate space $\tilde{\cal H}$,
so that we have to handle quantum mechanics on ${\cal H}\otimes\tilde{\cal H}$.
Let $a^{\dagger}$ and $a$ be creation and annihilation operators for original bosonic particles
acting on ${\cal H}$.
Similarly, let $\tilde{a}^{\dagger}$ and $\tilde{a}$ be creation and annihilation operators
for tilde bosonic particles acting on $\tilde{\cal H}$.
They obey the following commutation relations:
\begin{equation}
[a,a^{\dagger}]=[\tilde{a},\tilde{a}^{\dagger}]=1,
\quad\quad
[a,\tilde{a}]=[a,\tilde{a}^{\dagger}]=0.
\end{equation}

In the TFD formalism,
every state lying on ${\cal H}\otimes\tilde{\cal H}$ is invariant under the tilde conjugation.
The tilde conjugation rules are given as follows:
\begin{eqnarray}
(XY)\tilde{\;\;}
&=&
\tilde{X}\tilde{Y}, \nonumber \\
(\zeta_{1}X+\zeta_{2}Y)\tilde{\;\;}
&=&
\zeta_{1}^{*}\tilde{X}+\zeta_{2}^{*}\tilde{Y}, \nonumber \\
(X^{\dagger})\tilde{\;\;}
&=&
\tilde{X}^{\dagger}, \nonumber \\
(\tilde{X})\tilde{\;\;}
&=&
X, \nonumber \\
|0,\tilde{0}\rangle\tilde{\;\;}
&=&|0,\tilde{0}\rangle, \nonumber \\
\langle 0,\tilde{0}|\tilde{\;\;}
&=&\langle 0,\tilde{0}|,
\label{tilde-conjugation-1}
\end{eqnarray}
where $X$ and $Y$ are arbitrary bosonic operators acting on ${\cal H}$,
$\zeta_{1}$ and $\zeta_{2}$ are arbitrary complex numbers,
$\tilde{X}$ and $\tilde{Y}$ are tilde conjugates of $X$ and $Y$,
respectively.
Thus,
$\tilde{X}$ and $\tilde{Y}$ are operators that act on $\tilde{\cal H}$.
Moreover,
$|0,\tilde{0}\rangle=|0\rangle\otimes|\tilde{0}\rangle$
represents the direct product of $|0\rangle$ and $|\tilde{0}\rangle$,
which are vacuum states of an ordinary Hilbert space ${\cal H}$
and a tilde Hilbert space $\tilde{\cal H}$,
respectively.

Second,
we explain the Lie-Trotter product formula.
It is given as follows
\cite{Trotter1959,Kato1978,Reed1980}:
Let $A$ and $B$ be arbitrary Hermitian operators,
then
\begin{equation}
\lim_{N\to\infty}
(e^{itA/N}e^{itB/N})^{N}
=
e^{it(A+B)}
\quad\quad
\forall t>0.
\label{Lie-Trotter-product-formula-1}
\end{equation}
The Lie-Trotter product formula has a wide range of applications in theoretical physics.
It is useful for evaluating the Feynman path integral
because it divides the propagator into infinitesimal time evolution operators,
that is to say,
kinematic energy operators depending upon momentum only
and potential operators depending upon position only
\cite{Faris1967,Albeverio1976}.
It is also made use of for Monte Carlo simulations of quantum systems
because it gives tractable approximations to partition functions
\cite{Suzuki1976,Suzuki1993}.

\section{\label{section-definition-thermal-coherent-states}Definitions of thermal coherent states}
In this section,
using the Lie-Trotter product formula,
we define a new thermal coherent state.
Moreover,
we show that it is equivalent to conventional thermal coherent states except for different parameterizations and a phase factor.

At first,
we think about conventional thermal coherent states.
There are two natural candidates for these states,
\begin{eqnarray}
|\alpha,\zeta;\beta)
&=&
U(\beta)D(\alpha,\zeta)|0,\tilde{0}\rangle, \nonumber \\
|| \alpha,\zeta;\beta\rangle\!\rangle
&=&
D(\alpha,\zeta)U(\beta)|0,\tilde{0}\rangle,
\label{thermal-coherent-states-definition-old}
\end{eqnarray}
where
\begin{equation}
D(\alpha,\zeta)
=
\exp(\alpha a^{\dagger}-\alpha^{*}a
+\zeta\tilde{a}^{\dagger}-\zeta^{*}\tilde{a}),
\end{equation}
\begin{equation}
U(\beta)
=
e^{i\theta(\beta)G},
\end{equation}
\begin{equation}
G=i(a\tilde{a}-\tilde{a}^{\dagger}a^{\dagger}),
\end{equation}
\begin{eqnarray}
\cosh\theta(\beta)
&=&
(1-e^{-\beta\epsilon})^{-1/2}, \nonumber \\
\sinh\theta(\beta)
&=&
(e^{\beta\epsilon}-1)^{-1/2},
\label{finite-temperature-0}
\end{eqnarray}
\begin{equation}
\epsilon=\hbar\omega,
\label{planck-0}
\end{equation}
\begin{equation}
\beta
=
1/(k_{\mbox{\scriptsize B}}T).
\label{finite-temperature-1}
\end{equation}

In the current paper, we propose another thermal coherent state as follows:
\begin{eqnarray}
|\alpha,\zeta;\beta\rangle
&=&
\lim_{N\to\infty}
[U(\beta)^{1/N}D(\alpha,\zeta)^{1/N}]^{N}|0,\tilde{0}\rangle \nonumber \\
&=&
\exp[i\theta(\beta)G
+
(\alpha a^{\dagger}-\alpha^{*}a
+\zeta\tilde{a}^{\dagger}-\zeta^{*}\tilde{a})]|0,\tilde{0}\rangle.
\label{thermal-coherent-state-new-definition-1}
\end{eqnarray}
For the derivation of the above equation,
we use Eq.~(\ref{Lie-Trotter-product-formula-1}).
In the definitions of the conventional thermal coherent states
shown in Eq.~(\ref{thermal-coherent-states-definition-old}),
the thermalizing operator and the displacement operator are treated as individual processes.
By contrast,
in the definition of our thermal coherent state
shown in Eq.~(\ref{thermal-coherent-state-new-definition-1}),
both the thermalizing and displacement operators appear at the same time as a single process
and they are treated symmetrically.

These three thermal coherent states,
$|\alpha,\zeta;\beta)$, 
$|| \alpha,\zeta;\beta\rangle\!\rangle$ and
$|\alpha,\zeta;\beta\rangle$,
are essentially equivalent to each other.
In fact,
we can transform any one of them to the others with changing parameters $\alpha$ and $\zeta$
and multiplying the state vector by a proper phase factor.
In Appendix~\ref{section-Appendix-A},
we explain how to change $|\alpha,\zeta;\beta)$
into $|| \alpha,\zeta;\beta\rangle\!\rangle$
by replacing parameters $\alpha$ and $\zeta$.

Here,
we show that $|\alpha,\zeta;\beta\rangle$ given by Eq.~(\ref{thermal-coherent-state-new-definition-1})
is equivalent to $|\alpha,\zeta;\beta)$ given by Eq.~(\ref{thermal-coherent-states-definition-old}).
First,
we pay attention to the following relation:
\begin{eqnarray}
&&
[U(\theta/N)D(\alpha/N,\zeta/N)]^{n} \nonumber \\
&=&
\exp[(1/N^{2})\sum_{m=1}^{n-1}(n-m)\sinh(m\theta/N)(\alpha\zeta-\alpha^{*}\zeta^{*})]
U(n\theta/N) \nonumber \\
&&
\times
D((\alpha/N)\sum_{m=0}^{n-1}\cosh(m\theta/N)-(\zeta^{*}/N)\sum_{m=0}^{n-1}\sinh(m\theta/N), \nonumber \\
&&
\quad\quad
(\zeta/N)\sum_{m=0}^{n-1}\cosh(m\theta/N)-(\alpha^{*}/N)\sum_{m=0}^{n-1}\sinh(m\theta/N)) \nonumber \\
&&
\quad\quad\quad
\mbox{for $n=1,2,3,...$,}
\label{formula-induction-general-n}
\end{eqnarray}
where $U(\theta)=\exp(i\theta G)$.
We can derive Eq.~(\ref{formula-induction-general-n})
with the mathematical induction,
using Eqs.~(\ref{displacement-operator-formula-1}) and (\ref{formula-gamma-alpha-1})
in Appendix~\ref{section-Appendix-A}
and the Baker-Campbell-Hausdorff formula
\cite{Louisell1973,Walls1994}.
Second,
we evaluate the following limits of series:
\begin{eqnarray}
\lim_{N\to\infty}
\frac{1}{N^{2}}
\sum_{m=1}^{N-1}(N-m)\sinh\frac{m\theta}{N}
&=&
\frac{e^{\theta}-2\theta-e^{-\theta}}{2\theta^{2}}, \nonumber \\
\lim_{N\to\infty}
\frac{1}{N}
\sum_{m=0}^{N-1}\cosh\frac{m\theta}{N}
&=&
\frac{1}{\theta}\sinh\theta, \nonumber \\
\lim_{N\to\infty}
\frac{1}{N}
\sum_{m=0}^{N-1}\sinh\frac{m\theta}{N}
&=&
-\frac{1}{\theta}(1-\cosh\theta).
\end{eqnarray}

Finally,
from these results,
we achieve:
\begin{eqnarray}
|\alpha,\zeta;\beta\rangle
&=&
e^{i\Theta}
U(\beta)D(\alpha',\zeta')|0,\tilde{0}\rangle \nonumber \\
&=&
e^{i\Theta}|\alpha',\zeta';\beta),
\label{replacement-formula-thermal-coherent-states-A}
\end{eqnarray}
where
\begin{equation}
\Theta
=
-i\frac{e^{\theta}-2\theta-e^{-\theta}}{2\theta^{2}}(\alpha\zeta-\alpha^{*}\zeta^{*}),
\end{equation}
\begin{equation}
\left\{
\begin{array}{lll}
\alpha' & = & [\alpha\sinh\theta(\beta)+\zeta^{*}(1-\cosh\theta(\beta))]/\theta(\beta), \\
\zeta' & = & [\zeta\sinh\theta(\beta)+\alpha^{*}(1-\cosh\theta(\beta))]/\theta(\beta). \\
\end{array}
\right.
\label{replacement-formula-thermal-coherent-states-B}
\end{equation}
Now,
we pay attention to the following facts.
Because $(\alpha\zeta-\alpha^{*}\zeta^{*})$ is a pure imaginary number,
$\Theta$ is real.
Hence,
$e^{i\Theta}$ is just a phase factor,
and it never has effects on expectation values of physical quantities.
Thus,
we can conclude that $|\alpha,\zeta;\beta\rangle$ is essentially equivalent to $|\alpha',\zeta';\beta)$
with changing parameters $\alpha$ and $\zeta$
into $\alpha'$ and $\zeta'$ in accordance with Eq.~(\ref{replacement-formula-thermal-coherent-states-B}).

The criterion that we use to judge an arbitrary quantum state $|\psi(\beta)\rangle$
to be a thermal coherent state is expressed by:
\begin{equation}
\xi|\psi(\beta)\rangle
=
f(\beta)|\psi(\beta)\rangle,
\label{condition-eigenvector-xi-1}
\end{equation}
where
\begin{eqnarray}
\xi
&=&
U(\beta)a U^{\dagger}(\beta) \nonumber \\
&=&
\cosh\theta(\beta)a-\sinh\theta(\beta)\tilde{a}^{\dagger},
\label{definition-operator-xi-0}
\end{eqnarray}
and $f(\beta)$ has to be a c-number function.
[We can obtain Eq.~(\ref{definition-operator-xi-0})
using the Baker-Campbell-Hausdorff formula.]
We notice $|\alpha,\zeta;\beta)$ being an eigenvector of $\xi$
and its eigenvalue being equal to $\alpha$ with ease,
\begin{eqnarray}
\xi|\alpha,\zeta;\beta)
&=&
U(\beta)a U^{\dagger}(\beta)U(\beta)D(\alpha,\zeta)|0,\tilde{0}\rangle \nonumber \\
&=&
\alpha|\alpha,\zeta;\beta).
\label{eigenvector-0}
\end{eqnarray}
Thus,
from Eq.~(\ref{equivalence-old-thermal-coherent-states-1}) in Appendix~\ref{section-Appendix-A},
we realize $||\alpha,\zeta;\beta\rangle\!\rangle$ is an eigenvector of $\xi$ and its eigenvalue is equal to
$[\alpha\cosh\theta(\beta)-\zeta^{*}\sinh\theta(\beta)]$.
In a similar way,
from Eqs.~(\ref{replacement-formula-thermal-coherent-states-A}) and (\ref{replacement-formula-thermal-coherent-states-B}),
we realize $|\alpha,\zeta;\beta\rangle$ is an eigenvector of $\xi$ and its eigenvalue is equal to
$[\alpha\sinh\theta(\beta)+\zeta^{*}(1-\cosh\theta(\beta))]/\theta(\beta)$.

Here,
to gain a deeper understanding of the thermal coherent states,
we try constructing an eigenvector of $\xi$.
At first,
we consider a thermal vacuum state characterized by the parameter $\theta(\beta)$,
\begin{equation}
|0(\theta)\rangle
=
U(\beta)|0,\tilde{0}\rangle.
\end{equation}
Then,
the following relation holds,
\begin{eqnarray}
\xi|0(\theta)\rangle
&=&
U(\beta)a|0,\tilde{0}\rangle \nonumber \\
&=&
0.
\end{eqnarray}

Next,
using the Baker-Campbell-Hausdorff formula,
we obtain
\begin{equation}
e^{-f\xi^{\dagger}+f^{*}\xi}\xi e^{f\xi^{\dagger}-f^{*}\xi}
=
\xi+f,
\end{equation}
where $f$ is an arbitrary constant,
and
we can construct an eigenvector of $\xi$,
\begin{eqnarray}
|\phi\rangle
&=&
e^{f\xi^{\dagger}-f^{*}\xi}|0(\theta)\rangle \nonumber \\
&=&
e^{-(1/2)|f|^{2}}e^{f\xi^{\dagger}}|0(\theta)\rangle.
\label{eigenvector-xi-0}
\end{eqnarray}
In fact,
we can confirm that the state $|\phi\rangle$ given by Eq.~(\ref{eigenvector-xi-0}) is an eigenvector
of $\xi$ and its eigenvalue is equal to $f$ straightforwardly.

However,
the state $|\phi\rangle$ given by Eq.~(\ref{eigenvector-xi-0}) cannot be a proper thermal coherent state.
We examine this point in the following.
Using Eq.~(\ref{definition-operator-xi-0}),
we rewrite Eq.~(\ref{eigenvector-xi-0}) as:
\begin{eqnarray}
|\phi\rangle
&=&
\exp(\mu a^{\dagger}-\mu^{*}a
+
\nu\tilde{a}^{\dagger}-\nu^{*}\tilde{a})
|0(\theta)\rangle \nonumber \\
&=&
D(\mu,\nu)U(\beta)|0,\tilde{0}\rangle,
\label{eigenvector-xi-1}
\end{eqnarray}
where
\begin{eqnarray}
\mu&=&\cosh\theta(\beta)f, \nonumber \\
\nu&=&\sinh\theta(\beta)f^{*}.
\label{alpha_dash-0}
\end{eqnarray}

Now,
because of the invariance under the tilde conjugation for $|\phi\rangle$
given by Eqs.~(\ref{eigenvector-xi-0}), (\ref{eigenvector-xi-1}) and (\ref{alpha_dash-0}),
we obtain $\mu^{*}=\nu$,
that is to say,
\begin{equation}
\cosh\theta(\beta)f^{*}
=
\sinh\theta(\beta)f^{*}.
\label{eigenvector-xi-2}
\end{equation}
However,
Eq.~(\ref{eigenvector-xi-2}) does not hold for arbitrary $\theta(\beta)$.
This fact tells us that $|\phi\rangle$ given by Eq.~(\ref{eigenvector-xi-1}) is not a proper state
and we can hardly find a physical meaning from it.
From this discussion,
we understand that we cannot construct thermal coherent states in a simple manner as shown in Eq.~(\ref{eigenvector-xi-0}).
In other words,
an arbitrary eigenvector of $\xi$ is not always a thermal coherent state.

\section{\label{section-properties-thermal-coherent-state}An uncertainty relation
and quasiprobability distributions of $|\alpha,\zeta;\beta\rangle$}
In this section,
we examine an uncertainty relation and quasiprobability distributions of $|\alpha,\zeta;\beta\rangle$.

First,
as shown in Appendix~\ref{section-Appendix-A},
an uncertainty relation between momentum and position of $|\alpha,\zeta;\beta\rangle$ is given by:
\begin{eqnarray}
\Delta Q \Delta P
&=&
\frac{\hbar}{2}[\cosh^{2}\theta(\beta)+\sinh^{2}\theta(\beta)] \nonumber \\
&\geq&
\frac{\hbar}{2},
\label{uncertainity-relation-0}
\end{eqnarray}
where
\begin{eqnarray}
P
&=&
i\sqrt{\frac{\lambda\hbar}{2}}(a^{\dagger}-a), \nonumber \\
Q
&=&
\sqrt{\frac{\hbar}{2\lambda}}(a^{\dagger}+a),
\label{operator-position-momentum-0}
\end{eqnarray}
and uncertainty in an operator $A$ is defined as:
\begin{equation}
\Delta A=(\langle A^{2}\rangle-\langle A\rangle^{2})^{1/2}.
\label{definition-operator-uncertainty-0}
\end{equation}
[The uncertainty relation given by Eq.~(\ref{uncertainity-relation-0}) holds
for
$|\alpha,\zeta;\beta)$
and
$||\alpha,\zeta;\beta\rangle\!\rangle$,
as well.
These facts are shown in Appendix~\ref{section-Appendix-A}.]

Here,
we let the whole system be invariant under the tilde conjugation.
Thus,
we require the tilde conjugation defined by Eq.~(\ref{tilde-conjugation-1})
to leave $|\alpha,\zeta;\beta\rangle$, $|\alpha,\zeta;\beta)$ and $||\alpha,\zeta;\beta\rangle\!\rangle$
given by Eqs.~(\ref{thermal-coherent-states-definition-old})
and (\ref{thermal-coherent-state-new-definition-1})
invariant.
Hence,
we obtain the following relation between $\alpha$ and $\zeta$,
\begin{equation}
\alpha^{*}=\zeta.
\label{tilde-conjugation-2}
\end{equation}

\begin{figure}
\begin{center}
\includegraphics[scale=1.0]{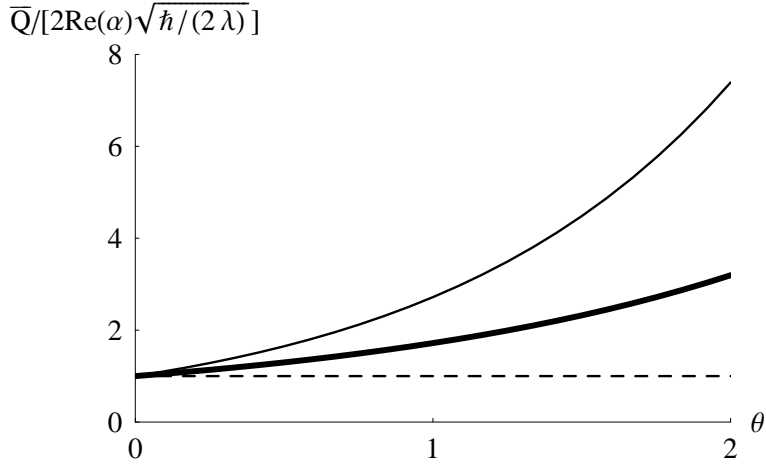}
\caption{Graphs of $\bar{Q}/[2\mbox{Re}(\alpha)\sqrt{\hbar/(2\lambda)}]$
against $\theta(\beta)$
for $|\alpha,\alpha^{*};\beta\rangle$, $|\alpha,\alpha^{*};\beta)$ and $||\alpha,\alpha^{*};\beta\rangle\!\rangle$,
where $\bar{Q}$ represents the expectation value of $Q$.
A thick solid curve, a thin solid curve, and a thin dashed line
stand for $|\alpha,\alpha^{*};\beta\rangle$, $|\alpha,\alpha^{*};\beta)$ and $||\alpha,\alpha^{*};\beta\rangle\!\rangle$,
respectively.}
\label{Figure01}
\end{center}
\end{figure}

Then,
we obtain expectation values of $Q$ and $P$
for $|\alpha,\alpha^{*};\beta\rangle$, $|\alpha,\alpha^{*};\beta)$ and $||\alpha,\alpha^{*};\beta\rangle\!\rangle$
as follows.
[The expectation values of $Q$ and $P$
for $|\alpha,\alpha^{*};\beta)$ and $||\alpha,\alpha^{*};\beta\rangle\!\rangle$
are evaluated in Appendix~\ref{section-Appendix-A}.]
We can compute the expectation values of $Q$ and $P$ for $|\alpha,\alpha^{*};\beta\rangle$ by replacing $\alpha$
of $(Q)$ and $(P)$ given by Eqs.~(\ref{expectation-value-Q-2}) and (\ref{expectation-value-P-2})
according to Eq.~(\ref{replacement-formula-thermal-coherent-states-B}).
Thinking about Eq.~(\ref{tilde-conjugation-2}),
we obtain the expectation values:
\begin{eqnarray}
\langle Q\rangle
&=&
\sqrt{\frac{\hbar}{2\lambda}}\frac{\cosh\theta(\beta)+\sinh\theta(\beta)-1}{\theta(\beta)}
(\alpha+\alpha^{*}), \nonumber \\
\langle P\rangle
&=&
i\sqrt{\frac{\lambda\hbar}{2}}\frac{\cosh\theta(\beta)+\sinh\theta(\beta)-1}{\theta(\beta)}
(\alpha^{*}-\alpha),
\end{eqnarray}
for $|\alpha,\alpha^{*};\beta\rangle$.
Similarly,
from Eqs.~(\ref{tilde-conjugation-2}), (\ref{expectation-value-Q-2}) and (\ref{expectation-value-P-2}),
we obtain
\begin{eqnarray}
(Q)
&=&
\sqrt{\frac{\hbar}{2\lambda}}[\cosh\theta(\beta)+\sinh\theta(\beta)]
(\alpha+\alpha^{*}), \nonumber \\
(P)
&=&
i\sqrt{\frac{\lambda\hbar}{2}}[\cosh\theta(\beta)+\sinh\theta(\beta)]
(\alpha^{*}-\alpha),
\end{eqnarray}
for $|\alpha,\alpha^{*};\beta)$.
Furthermore,
from Eqs.~(\ref{tilde-conjugation-2}), (\ref{expectation-value-Q-3}) and (\ref{expectation-value-P-3}),
we obtain:
\begin{eqnarray}
\langle\!\langle Q\rangle\!\rangle
&=&
\sqrt{\frac{\hbar}{2\lambda}}
(\alpha+\alpha^{*}), \nonumber \\
\langle\!\langle P\rangle\!\rangle
&=&
i\sqrt{\frac{\lambda\hbar}{2}}
(\alpha^{*}-\alpha),
\end{eqnarray}
for $||\alpha,\alpha^{*};\beta\rangle\!\rangle$.
In Fig.~\ref{Figure01},
we plot $\bar{Q}/[2\mbox{Re}(\alpha)\sqrt{\hbar/(2\lambda)}]$ against $\theta(\beta)$
for
$|\alpha,\alpha^{*};\beta\rangle$, $|\alpha,\alpha^{*};\beta)$ and $||\alpha,\alpha^{*};\beta\rangle\!\rangle$
with a thick solid curve, a thin solid curve, and a thin dashed line,
respectively,
where $\bar{Q}$ represents the expectation value of $Q$.

Second,
we derive the $P$-representation of
$|\alpha,\alpha^{*};\beta\rangle$.
The $P$-representations of $|\alpha,\alpha^{*};\beta)$ and $||\alpha,\alpha^{*};\beta\rangle\!\rangle$
are derived in Appendix~\ref{section-P-representation-conventional}.
Replacing $\alpha$ of $P(\mu)$ for $|\alpha,\alpha^{*};\beta)$ given by Eq.~(\ref{P-representation-final-2})
according to Eq.~(\ref{replacement-formula-thermal-coherent-states-B})
with Eq.~(\ref{tilde-conjugation-2}),
we obtain:
\begin{equation}
P(\mu)
=
G(\mu;
\frac{\sinh\theta(\beta)+\cosh\theta(\beta)-1}{\theta(\beta)}\alpha,
\frac{1}{\sqrt{2}}\sinh\theta(\beta)),
\label{P-representation-final-1}
\end{equation}
where
\begin{equation}
G(x;\bar{x},\sigma)
=
\frac{1}{2\pi\sigma^{2}}
\exp(-\frac{|x-\bar{x}|^{2}}{2\sigma^{2}}).
\label{Gaussian-function}
\end{equation}

\begin{figure}
\begin{center}
\includegraphics[scale=1.0]{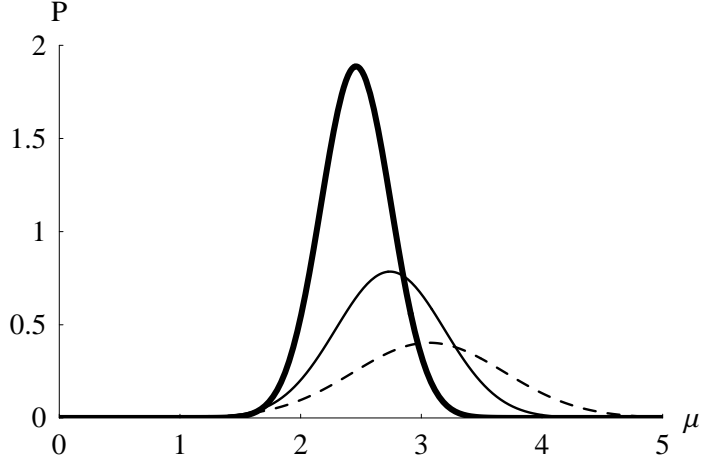}
\caption{Graphs of $P(\mu)$ of $|\alpha,\alpha^{*};\beta\rangle$ given by Eq.~(\ref{P-representation-final-1})
against $\mu$ with $\alpha=2.0$ are plotted.
A thick solid curve, a thin solid curve, and a thin dashed curve represent graphs of
$\theta=0.4$, $\theta=0.6$ and $\theta=0.8$,
respectively.}
\label{Figure02}
\end{center}
\end{figure}

\begin{figure}
\begin{center}
\includegraphics[scale=1.0]{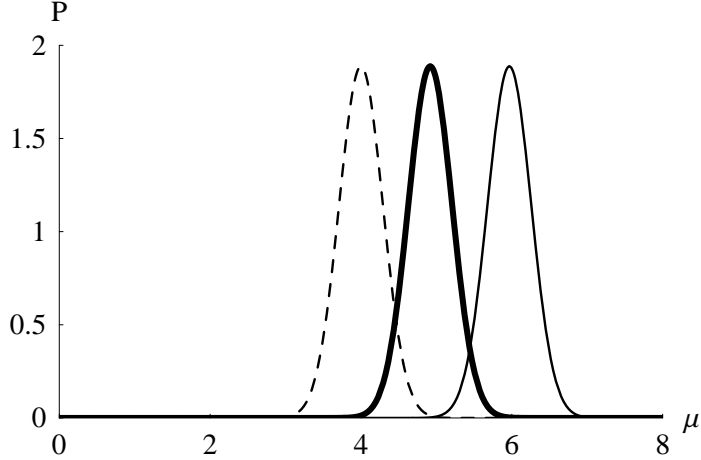}
\caption{Graphs of $P(\mu)$ against $\mu$
for
$|\alpha,\alpha^{*};\beta\rangle$, $|\alpha,\alpha^{*};\beta)$ and $||\alpha,\alpha^{*};\beta\rangle\!\rangle$
are plotted with
a thick solid curve, a thin solid curve, and a thin dashed curve,
respectively.
Setting $\alpha=4.0$ and $\theta=0.4$,
we draw curves of $P(\mu)$ given by
Eqs.~(\ref{P-representation-final-1}), (\ref{P-representation-final-2})
and (\ref{P-representation-final-3}).}
\label{Figure03}
\end{center}
\end{figure}

In Fig.~\ref{Figure02},
graphs of $P(\mu)$ for $|\alpha,\alpha^{*};\beta\rangle$ given by Eq.~(\ref{P-representation-final-1})
against $\mu$ with $\alpha=2.0$ are plotted.
A thick solid curve, a thin solid curve, and a thin dashed curve represent graphs of
$\theta=0.4$, $\theta=0.6$ and $\theta=0.8$,
respectively.
In Fig.~\ref{Figure03},
we plot graphs of $P(\mu)$ against $\mu$
for $|\alpha,\alpha^{*};\beta\rangle$, $|\alpha,\alpha^{*};\beta)$ and $||\alpha,\alpha^{*};\beta\rangle\!\rangle$
with
a thick solid curve, a thin solid curve, and a thin dashed curve,
respectively.
Setting $\alpha=4.0$ and $\theta=0.4$,
we draw curves of $P(\mu)$ given by
Eqs.~(\ref{P-representation-final-1}), (\ref{P-representation-final-2})
and (\ref{P-representation-final-3}) in Fig.~\ref{Figure03}.

Third,
we derive the $Q$-function of $|\alpha,\alpha^{*};\beta\rangle$.
The $Q$-functions of $|\alpha,\alpha^{*};\beta)$ and $||\alpha,\alpha^{*},\beta\rangle\!\rangle$
are derived in Appendix~\ref{section-Q-representation-conventional}.
Applying a replacement of Eq.~(\ref{replacement-formula-thermal-coherent-states-B})
with Eq.~(\ref{tilde-conjugation-2}) to $Q(\mu)$ for $|\alpha,\alpha^{*};\beta)$ given by Eq.~(\ref{Q-representation-final-2}),
we obtain:
\begin{equation}
Q(\mu)
=
G(\mu;
\frac{\sinh\theta(\beta)+\cosh\theta(\beta)-1}{\theta(\beta)}\alpha,
\frac{1}{\sqrt{2}}\cosh\theta(\beta)),
\label{Q-representation-final-1}
\end{equation}
where $G(x;\bar{x},\sigma)$ is given by Eq.~(\ref{Gaussian-function}).

Fourth,
we consider the Wigner function of $|\alpha,\alpha^{*};\beta\rangle$.
The Wigner functions of $|\alpha,\alpha^{*};\beta)$ and $||\alpha,\alpha^{*},\beta\rangle\!\rangle$
are derived in Appendix~\ref{section-Q-representation-conventional}.
Applying a replacement of Eq.~(\ref{replacement-formula-thermal-coherent-states-B})
with Eq.~(\ref{tilde-conjugation-2}) to $W(\mu)$ for $|\alpha,\alpha^{*};\beta)$ given by Eq.~(\ref{W-representation-final-2}),
we can compute $W(\mu)$ as
\begin{equation}
W(\mu)
=
G(\mu;
\frac{\sinh\theta(\beta)+\cosh\theta(\beta)-1}{\theta(\beta)}\alpha,
\frac{1}{2}[\cosh^{2}\theta(\beta)+\sinh^{2}\theta(\beta)]^{1/2}).
\label{W-representation-final-1}
\end{equation}

\section{\label{section-OPO-laser}Experimental realization of
$|\gamma_{\mbox{\scriptsize s}},\gamma_{\mbox{\scriptsize i}};\beta\rangle$
with the OPO laser}
In this section,
we discuss an experimental setup of the OPO laser system
for realizing $|\gamma_{\mbox{\scriptsize s}},\gamma_{\mbox{\scriptsize i}};\beta\rangle$
actually in the laboratory,
where $\gamma_{\mbox{\scriptsize s}}$ and $\gamma_{\mbox{\scriptsize i}}$
represent constants of displacement operators of the signal and idler photons,
respectively.
Pumped by a single-frequency laser,
the OPO system generates both the signal light and the idler light
as continuously tunable frequency radiations.
Thus, the OPO system is suitable for nonlinear frequency conversion
\cite{Grynberg2010,Michel1998a}.

\begin{figure}
\begin{center}
\includegraphics[scale=1.2]{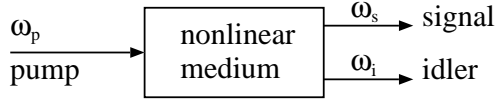}
\caption{A schematic diagram of the OPO.}
\label{Figure04}
\end{center}
\end{figure}

The OPO is based on a nonlinear optical process that involves energy conservation of a pump beam
at fixed frequency of $\omega_{\mbox{\scriptsize p}}$,
into two lower energy beams,
the signal at $\omega_{\mbox{\scriptsize s}}$ and the idler at $\omega_{\mbox{\scriptsize i}}$.
Figure~\ref{Figure04} shows a schematic diagram of the OPO.
The $\beta$-barium borate (BBO) is commonly used as a nonlinear optical crystal.
The lithium triborate can also be made use of as a nonlinear medium.
We can utilize the Nd:YAG laser for pumping the OPO.

In the OPO,
a pump photon with a wave vector $\mbox{\boldmath $k$}_{\mbox{\scriptsize p}}$ at a circular frequency $\omega_{\mbox{\scriptsize p}}$
breaks down into two lower frequency signal and idler photons with wave vectors
$\mbox{\boldmath $k$}_{\mbox{\scriptsize s}}$ and $\mbox{\boldmath $k$}_{\mbox{\scriptsize i}}$
at $\omega_{\mbox{\scriptsize s}}$ and $\omega_{\mbox{\scriptsize i}}$,
respectively.
During this process,
the energy conservation is satisfied as
$\omega_{\mbox{\scriptsize p}}=\omega_{\mbox{\scriptsize s}}+\omega_{\mbox{\scriptsize i}}$.
At the same time,
the momentum conservation is expressed as
$\mbox{\boldmath $k$}_{\mbox{\scriptsize p}}=\mbox{\boldmath $k$}_{\mbox{\scriptsize s}}+\mbox{\boldmath $k$}_{\mbox{\scriptsize i}}$.
The norm of the wave vector $\mbox{\boldmath $k$}$ is given by
$|\mbox{\boldmath $k$}|=n\omega/c$,
where $n$ is the refractive index and $c$ is the speed of light in vacuum.
Thus,
for example,
if $\mbox{\boldmath $k$}_{\mbox{\scriptsize p}}$, $\mbox{\boldmath $k$}_{\mbox{\scriptsize s}}$
and $\mbox{\boldmath $k$}_{\mbox{\scriptsize i}}$
are parallel to each other,
the following relation has to hold:
$n_{\mbox{\scriptsize p}}\omega_{\mbox{\scriptsize p}}
=
n_{\mbox{\scriptsize s}}\omega_{\mbox{\scriptsize s}}+n_{\mbox{\scriptsize i}}\omega_{\mbox{\scriptsize i}}$.
However,
dispersion in a nonlinear medium lets $n_{\mbox{\scriptsize p}}$,
$n_{\mbox{\scriptsize s}}$ and $n_{\mbox{\scriptsize i}}$ be different from each other,
so that
$n_{\mbox{\scriptsize p}}\omega_{\mbox{\scriptsize p}}
\neq
n_{\mbox{\scriptsize s}}\omega_{\mbox{\scriptsize s}}
+
n_{\mbox{\scriptsize i}}\omega_{\mbox{\scriptsize i}}$
in general.
To overcome this trouble,
we can use birefringence for phase matching.

\begin{figure}
\begin{center}
\includegraphics[scale=1.2]{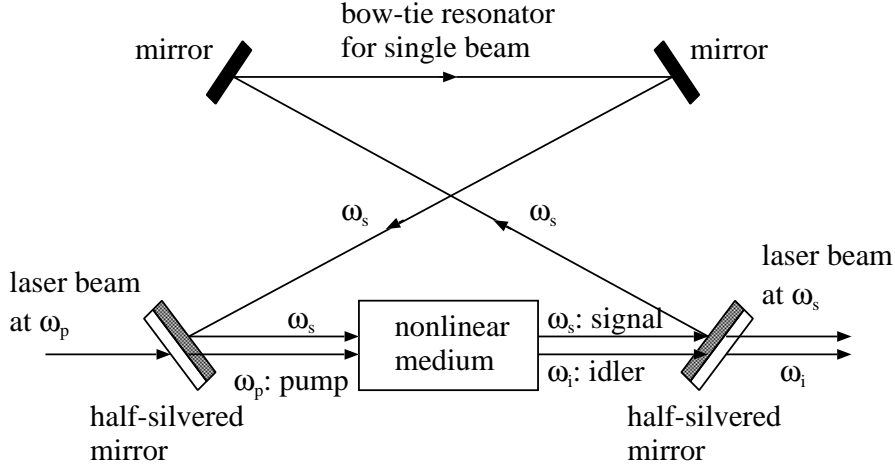}
\caption{The OPO-based laser system with single resonance
implemented by a bow-tie-type ring cavity.
It outputs a laser beam at $\omega_{\mbox{\scriptsize s}}$.}
\label{Figure05}
\end{center}
\end{figure}

\begin{figure}
\begin{center}
\includegraphics[scale=1.2]{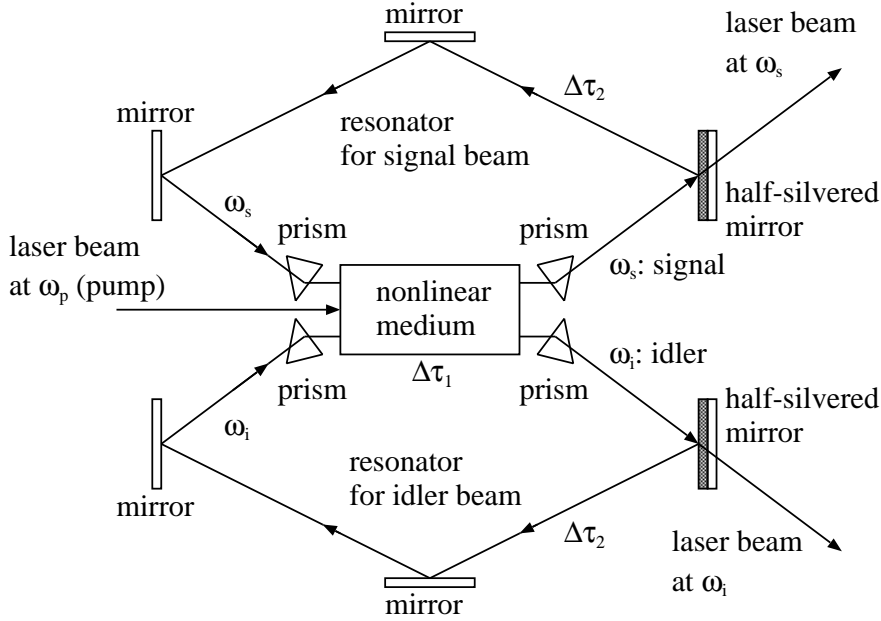}
\caption{The OPO-based laser system with double resonance.
It outputs two laser beams at $\omega_{\mbox{\scriptsize s}}$ and $\omega_{\mbox{\scriptsize i}}$.
The signal and idler beams travel through the crystal
for a period $\Delta\tau_{1}$ and through the resonators for a period $\Delta\tau_{2}$.
The signal and idler photons go round inside the OPO with the resonators $N$ times and their time of flight is equal to $T$ in total.}
\label{Figure06}
\end{center}
\end{figure}

The OPO-based laser system is constructed out of the OPO with ring resonators
as shown in Figs.~\ref{Figure05} and \ref{Figure06}
\cite{Michel1998a,Bosenberg1993,Gloster1995,Johnson1995}.
In Fig.~\ref{Figure05}, single resonance is caused in the bow-tie ring resonator,
in which only the signal radiation passes through and builds up in intensity
owing to constructive interference.
(This type of the OPO is examined
in Refs.~\cite{Suzuki2006,Arslanov2011}.)
If we put signal and idler beams into their ring resonators individually
in Fig.~\ref{Figure06},
we obtain double resonance and the system outputs laser beams
at $\omega_{\mbox{\scriptsize s}}$ and $\omega_{\mbox{\scriptsize i}}$.

In contrast to conventional lasers,
the OPO-based laser never induces population inversion in the parametric conversion process,
because it does not depend on an atomic or molecular transition.
The OPO laser system made out of the BBO crystal is explored for use as an excitation source
for laser-excited atomic fluorescence spectrometry
\cite{Michel1997,Michel1998b,Michel2001}.

Here, let us consider the Hamiltonian that describes the OPO-based laser system.
We can regard the OPO system as a nondegenerate parametric amplifier.
We assume that the amplitude of the pump mode is sufficiently strong and its loss can be neglected
during the nonlinear interaction.
Thus,
we treat the pump mode as classical light.
Moreover,
we assume that the phase-matching condition is fulfilled as
$\mbox{\boldmath $k$}_{\mbox{\scriptsize p}}=\mbox{\boldmath $k$}_{\mbox{\scriptsize s}}+\mbox{\boldmath $k$}_{\mbox{\scriptsize i}}$.

The classical pump mode at frequency $\omega_{\mbox{\scriptsize p}}$ interacts in a nonlinear optical medium
with signal and idler modes at frequencies $\omega_{\mbox{\scriptsize s}}$ and $\omega_{\mbox{\scriptsize i}}$, respectively.
Because of the conservation of the energy,
these frequencies satisfy the relation $\omega_{\mbox{\scriptsize p}}=\omega_{\mbox{\scriptsize s}}+\omega_{\mbox{\scriptsize i}}$.

The Hamiltonian of the OPO system is given by:
\begin{equation}
H
=
\hbar\omega_{\mbox{\scriptsize s}}a_{\mbox{\scriptsize s}}^{\dagger}a_{\mbox{\scriptsize s}}
+
\hbar\omega_{\mbox{\scriptsize i}}a_{\mbox{\scriptsize i}}^{\dagger}a_{\mbox{\scriptsize i}}
+
i\hbar\chi^{(2)}
(a_{\mbox{\scriptsize s}}^{\dagger}a_{\mbox{\scriptsize i}}^{\dagger}e^{-2i\omega_{\mbox{\scriptsize p}}t}
-a_{\mbox{\scriptsize s}}a_{\mbox{\scriptsize i}}e^{2i\omega_{\mbox{\scriptsize p}}t}),
\end{equation}
where $a_{\mbox{\scriptsize s}}$ and $a_{\mbox{\scriptsize i}}$ are the annihilation operators for the signal and idler modes,
respectively
\cite{Walls1994}.
The coupling constant $\chi^{(2)}$ is proportional
to the second-order susceptibility of the medium
and to the amplitude of the pump.
Taking the interaction picture,
we obtain the time-independent Hamiltonian,
\begin{equation}
H_{\mbox{\scriptsize I}}
=
i\hbar\chi^{(2)}
(a_{\mbox{\scriptsize s}}^{\dagger}a_{\mbox{\scriptsize i}}^{\dagger}-a_{\mbox{\scriptsize s}}a_{\mbox{\scriptsize i}}).
\label{nondegenerate-squeezing-Hamiltonian}
\end{equation}

Considering the double resonance,
we prepare another Hamiltonian,
which causes evolution of coherence of radiations at $\omega_{\mbox{\scriptsize s}}$ and $\omega_{\mbox{\scriptsize i}}$,
as follows:
\begin{equation}
H_{\mbox{\scriptsize II}}
=
i\hbar(g_{\mbox{\scriptsize s}}a_{\mbox{\scriptsize s}}^{\dagger}-g_{\mbox{\scriptsize s}}^{*}a_{\mbox{\scriptsize s}})
+
i\hbar(g_{\mbox{\scriptsize i}}a_{\mbox{\scriptsize i}}^{\dagger}-g_{\mbox{\scriptsize i}}^{*}a_{\mbox{\scriptsize i}}).
\end{equation}

Here,
we assume that the signal and idler beams travel through the crystal for a period $\Delta\tau_{1}$
and through the resonators for a period $\Delta\tau_{2}$
in the setup of Fig.~\ref{Figure06}.
Moreover,
we assume that the signal and idler photons go round inside the OPO with the resonators $N$ times
and their time of flight is equal to $T$ in total.
Thus,
the following relations hold:
\begin{eqnarray}
T&=&T_{1}+T_{2}, \nonumber \\
T_{1}&=&N\Delta\tau_{1}, \nonumber \\
T_{2}&=&N\Delta\tau_{2}.
\end{eqnarray}

Then,
under the limit of $N\gg 1$ and on the assumption that $T_{1}$ and $T_{2}$ are put on certain fixed values,
we can write down the time evolution unitary operator as:
\begin{eqnarray}
U
&=&
\exp[-i\frac{\Delta\tau_{1}}{\hbar}H_{\mbox{\scriptsize I}}]
\exp[-i\frac{\Delta\tau_{2}}{\hbar}H_{\mbox{\scriptsize II}}]
...
\exp[-i\frac{\Delta\tau_{1}}{\hbar}H_{\mbox{\scriptsize I}}]
\exp[-i\frac{\Delta\tau_{2}}{\hbar}H_{\mbox{\scriptsize II}}] \nonumber \\
&=&
\Biggl(
\exp[-i\frac{T_{1}}{\hbar N}H_{\mbox{\scriptsize I}}]
\exp[-i\frac{T_{2}}{\hbar N}H_{\mbox{\scriptsize II}}]
\Biggr)^{N} \nonumber \\
&\simeq&
\exp[-i\frac{T_{1}}{\hbar}H_{\mbox{\scriptsize I}}-i\frac{T_{2}}{\hbar}H_{\mbox{\scriptsize II}}] \nonumber \\
&=&
\exp
[-\theta(a_{\mbox{\scriptsize s}}a_{\mbox{\scriptsize i}}-a_{\mbox{\scriptsize s}}^{\dagger}a_{\mbox{\scriptsize i}}^{\dagger})
+\gamma_{\mbox{\scriptsize s}}a_{\mbox{\scriptsize s}}^{\dagger}-\gamma_{\mbox{\scriptsize s}}^{*}a_{\mbox{\scriptsize s}}
+\gamma_{\mbox{\scriptsize i}}a_{\mbox{\scriptsize i}}^{\dagger}-\gamma_{\mbox{\scriptsize i}}^{*}a_{\mbox{\scriptsize i}}],
\label{unitary-operator-OPO}
\end{eqnarray}
where we use Eq.~(\ref{Lie-Trotter-product-formula-1}) for the above approximation,
$\theta=\chi^{(2)}T_{1}$,
$\gamma_{\mbox{\scriptsize s}}=g_{\mbox{\scriptsize s}}T_{2}$
and
$\gamma_{\mbox{\scriptsize i}}=g_{\mbox{\scriptsize i}}T_{2}$.
This unitary operator coincides with that of Eq.~(\ref{thermal-coherent-state-new-definition-1}),
which acts on the vacuum state $|0,\tilde{0}\rangle$
and transforms it into $|\alpha,\zeta;\beta\rangle$.
Thus,
generating the state of the whole system with the unitary transformation $U$
given by Eq.~(\ref{unitary-operator-OPO}),
and tracing out the degree of freedom of the idler photons,
we obtain the density operator of the signal photons
$\rho=\mbox{Tr}_{\mbox{\scriptsize idler}}
(|\gamma_{\mbox{\scriptsize s}},\gamma_{\mbox{\scriptsize i}};\beta\rangle
\langle\gamma_{\mbox{\scriptsize s}},\gamma_{\mbox{\scriptsize i}};\beta|)$.

Finally,
we indicate the following attention.
In the TFD,
the fictitious tilde particles induce the thermal dissipation for the original particles.
Contrastingly,
in the OPO-based laser system,
the idler photons cause the thermal fluctuation of the signal light.
If we put this difference aside,
we can consider the state of the OPO laser system to be equivalent to the thermal coherent state.

\section{\label{section-discussions}Discussions}
In this paper,
we investigate properties of the thermal coherent state defined with the Lie-Trotter product formula.
To examine the time evolution of a quantum system at finite temperature,
we can use the thermal coherent state for an initial state of radiation.
Thus,
it is important to prepare a wide variety of thermal coherent states
from both theoretical and experimental viewpoints.

Studying quantum statistical mechanics is a work in progress.
In particular,
methods for examining its nonequilibrium cases are not established yet.
Thus,
the authors think that to investigate a wide variety of thermal coherent states
can be a milestone in the development of quantum statistical mechanics.

In Sec.~\ref{section-OPO-laser},
we point out that our thermal coherent state is realized by the OPO-based laser system.
Therefore,
our thermal coherent state gives a description of an actual experiment.
This is one of the advantages that our thermal coherent state owns.
According to the consideration given in Sec.~\ref{section-OPO-laser},
the output of the signal light of the OPO is a thermal coherent state
and it can never be a coherent state at zero temperature.
Because of fictitious thermal dissipation given rise to by the Hamiltonian $H_{\mbox{\scriptsize I}}$
in Eq.~(\ref{nondegenerate-squeezing-Hamiltonian}),
output power of lasers based on optical parametric wavelength conversion
becomes weak.
However,
we can let Nd:YAG pump source be strong enough
for overcoming this trouble.

\appendix
\section{\label{section-Appendix-A}Equivalence of $|\alpha,\zeta;\beta)$ and $||\alpha,\zeta;\beta\rangle\!\rangle$,
their characteristic functions and uncertainty relations}
In this section,
first,
we show equivalence of $|\alpha,\zeta;\beta)$ and $||\alpha,\zeta;\beta\rangle\!\rangle$
given by Eq.~(\ref{thermal-coherent-states-definition-old}).
Next,
we derive their characteristic functions
and uncertainty relations between physical quantities,
that is to say,
momentum and position.

From Eq.~(\ref{thermal-coherent-states-definition-old}),
we can rewrite $||\alpha,\zeta;\beta\rangle\!\rangle$ as:
\begin{equation}
||\alpha,\zeta;\beta\rangle\!\rangle
=
U(\beta)U^{\dagger}(\beta)D(\alpha,\zeta)U(\beta)|0,\tilde{0}\rangle.
\label{another-form-thermal-coherent-state-dash0}
\end{equation}
Because of
\begin{eqnarray}
[a\tilde{a}-\tilde{a}^{\dagger}a^{\dagger},
\alpha a^{\dagger}-\alpha^{*}a+\zeta\tilde{a}^{\dagger}-\zeta^{*}\tilde{a}]
&=&
-(\zeta^{*} a^{\dagger}-\zeta a+\alpha^{*}\tilde{a}^{\dagger}-\alpha\tilde{a}),
\nonumber \\
{[}a\tilde{a}-\tilde{a}^{\dagger}a^{\dagger},
{[}a\tilde{a}-\tilde{a}^{\dagger}a^{\dagger},
\alpha a^{\dagger}-\alpha^{*}a+\zeta\tilde{a}^{\dagger}-\zeta^{*}\tilde{a}]]
&=&
\alpha a^{\dagger}-\alpha^{*}a+\zeta\tilde{a}^{\dagger}-\zeta^{*}\tilde{a},
\nonumber \\
&&\quad\cdots,
\end{eqnarray}
and using the Baker-Campbell-Hausdorff formula \cite{Louisell1973,Walls1994},
we obtain:
\begin{eqnarray}
&&
e^{-i\theta(\beta)G}
(\alpha a^{\dagger}-\alpha^{*}a+\zeta\tilde{a}^{\dagger}-\zeta^{*}\tilde{a})
e^{i\theta(\beta)G} \nonumber \\
&=&
[\alpha \cosh\theta(\beta)-\zeta^{*}\sinh\theta(\beta)]a^{\dagger}
-
[\alpha^{*} \cosh\theta(\beta)-\zeta\sinh\theta(\beta)]a \nonumber \\
&&
+
[\zeta\cosh\theta(\beta)-\alpha^{*}\sinh\theta(\beta)]\tilde{a}^{\dagger}
-
[\zeta^{*} \cosh\theta(\beta)-\alpha \sinh\theta(\beta)]\tilde{a}.
\end{eqnarray}

Thus,
we obtain:
\begin{eqnarray}
U^{\dagger}(\beta)D(\alpha,\zeta)U(\beta)
&=&
\exp[e^{-i\theta(\beta)G}
(\alpha a^{\dagger}-\alpha^{*}a+\zeta\tilde{a}^{\dagger}-\zeta^{*}\tilde{a})
e^{i\theta(\beta)G}] \nonumber \\
&=&
D(\alpha \cosh\theta(\beta)-\zeta^{*}\sinh\theta(\beta),
\zeta\cosh\theta(\beta)-\alpha^{*}\sinh\theta(\beta)).
\label{displacement-operator-formula-1}
\end{eqnarray}
From Eqs.~(\ref{thermal-coherent-states-definition-old}),
(\ref{another-form-thermal-coherent-state-dash0})
and (\ref{displacement-operator-formula-1}),
we arrive at:
\begin{eqnarray}
||\alpha,\zeta;\beta\rangle\!\rangle
&=&
U(\beta)
D(\alpha\cosh\theta(\beta)-\zeta^{*}\sinh\theta(\beta),
\zeta\cosh\theta(\beta)-\alpha^{*}\sinh\theta(\beta))|0,\tilde{0}\rangle \nonumber \\
&=&
|\alpha\cosh\theta(\beta)-\zeta^{*}\sinh\theta(\beta),\zeta\cosh\theta(\beta)-\alpha^{*}\sinh\theta(\beta);\beta).
\label{equivalence-old-thermal-coherent-states-1}
\end{eqnarray}
Looking at Eq.~(\ref{equivalence-old-thermal-coherent-states-1}),
we notice the following.
Replacing $\alpha$ and $\zeta$ of $|\alpha,\zeta;\beta)$ according to
\begin{equation}
\left\{
\begin{array}{lll}
\alpha & \longrightarrow & \alpha \cosh\theta(\beta)-\zeta^{*}\sinh\theta(\beta), \\
\zeta & \longrightarrow  & \zeta\cosh\theta(\beta)-\alpha^{*}\sinh\theta(\beta),
\end{array}
\right.
\label{alpha-replacement-1}
\end{equation}
we obtain $||\alpha,\zeta;\beta\rangle\!\rangle$,
so that $|\alpha,\zeta;\beta)$ and $||\alpha,\zeta;\beta\rangle\!\rangle$ are essentially equivalent to each other.

Next,
we compute the characteristic function of $|\alpha,\zeta;\beta)$.
We define the characteristic function of $|\alpha,\zeta;\beta)$ as follows:
\begin{equation}
(\mbox{CF})
=
(\alpha,\zeta;\beta|\exp[-i(qQ+pP+\tilde{q}\tilde{Q}+\tilde{p}\tilde{P})]|\alpha,\zeta;\beta),
\label{definition-characteristic-function-01}
\end{equation}
where $P$ and $Q$ are given by Eq.~(\ref{operator-position-momentum-0}) and
\begin{eqnarray}
\tilde{P}
&=&
-i\sqrt{\frac{\lambda\hbar}{2}}(\tilde{a}^{\dagger}-\tilde{a}), \nonumber \\
\tilde{Q}
&=&
\sqrt{\frac{\hbar}{2\lambda}}(\tilde{a}^{\dagger}+\tilde{a}).
\label{definition-characteristic-function-02}
\end{eqnarray}
Then,
letting $m$ and $\omega$ be the mass and the angular frequency of the harmonic oscillator,
respectively,
and assuming the relation $\lambda=m\omega$,
we obtain:
\begin{equation}
-i(qQ+pP+q'\tilde{Q}+p'\tilde{P})
=
\gamma a^{\dagger}-\gamma^{*}a+\gamma'\tilde{a}^{\dagger}-\gamma'^{*}\tilde{a},
\label{definition-characteristic-function-03}
\end{equation}
where
\begin{eqnarray}
\gamma
&=&
-iq\sqrt{\frac{\hbar}{2\lambda}}+p\sqrt{\frac{\lambda\hbar}{2}}, \nonumber \\
\gamma'
&=&
-iq'\sqrt{\frac{\hbar}{2\lambda}}-p'\sqrt{\frac{\lambda\hbar}{2}}.
\label{definition-characteristic-function-04}
\end{eqnarray}
Thus,
we obtain the displacement operator:
\begin{equation}
\exp[-i(qQ+pP+q'\tilde{Q}+p'\tilde{P})]
=
D(\gamma,\gamma').
\label{definition-characteristic-function-05}
\end{equation}
From Eqs.~(\ref{displacement-operator-formula-1}),
(\ref{definition-characteristic-function-01}) and (\ref{definition-characteristic-function-05}), we obtain
\begin{eqnarray}
(\mbox{CF})
&=&
(\alpha,\zeta;\beta|D(\gamma,\gamma')|\alpha,\zeta;\beta) \nonumber \\
&=&
\langle 0,\tilde{0}|D^{\dagger}(\alpha,\zeta)
D(\gamma\cosh\theta(\beta)-\gamma'^{*}\sinh\theta(\beta),
\gamma'\cosh\theta(\beta)-\gamma^{*}\sinh\theta(\beta)) \nonumber \\
&&
\times
D(\alpha,\zeta)|0,\tilde{0}\rangle.
\label{characteristic-function-2}
\end{eqnarray}

Here,
we prepare the following commutation relation:
\begin{equation}
[\gamma a^{\dagger}-\gamma^{*}a+\gamma'\tilde{a}^{\dagger}-\gamma'^{*}\tilde{a},
\alpha a^{\dagger}-\alpha^{*}a+\zeta\tilde{a}^{\dagger}-\zeta^{*}\tilde{a}]
=
\gamma \alpha^{*}-\gamma^{*}\alpha
+
\gamma'\zeta^{*}-\gamma'^{*}\zeta.
\label{formula-gamma-alpha-1}
\end{equation}
Because the right-hand side of Eq.~(\ref{formula-gamma-alpha-1}) is given by a c-number,
using the Baker-Campbell-Hausdorff formula,
we obtain:
\begin{equation}
D(\gamma,\gamma')D(\alpha,\zeta)
=
\exp
(\gamma \alpha^{*}-\gamma^{*}\alpha
+
\gamma'\zeta^{*}-\gamma'^{*}\zeta)
D(\alpha,\zeta)D(\gamma,\gamma').
\end{equation}
Thus,
we arrive at:
\begin{eqnarray}
(\mbox{CF})
&=&
\exp
\Biggl(
-\frac{1}{2}
[(\cosh^{2}\theta(\beta)+\sinh^{2}\theta(\beta))(|\gamma|^{2}+|\gamma'|^{2}) \nonumber \\
&&
-2\cosh\theta(\beta)\sinh\theta(\beta)(\gamma\gamma'+\gamma^{*}\gamma'^{*})] \nonumber \\
&&
+
(\gamma \cosh\theta(\beta)-\gamma'^{*}\sinh\theta(\beta))\alpha^{*}
-
(\gamma^{*} \cosh\theta(\beta)-\gamma'\sinh\theta(\beta))\alpha \nonumber \\
&&
+
(\gamma'\cosh\theta(\beta)-\gamma^{*}\sinh\theta(\beta))\zeta^{*}
-
(\gamma'^{*} \cosh\theta(\beta)-\gamma \sinh\theta(\beta))\zeta
\Biggr).
\label{characteristic-function-B}
\end{eqnarray}
From Eq.~(\ref{characteristic-function-B}),
the expectation values of $Q$ and $Q^{2}$ are given by:
\begin{eqnarray}
(Q)
&=&
\left.
i\frac{\partial}{\partial q}(\mbox{CF})
\right|_{\gamma=\gamma^{*}=\gamma'=\gamma'^{*}=0} \nonumber \\
&=&
\sqrt{\frac{\hbar}{2\lambda}}
[\cosh\theta(\beta)\alpha^{*}+\sinh\theta(\beta)\zeta
+\cosh\theta(\beta)\alpha+\sinh\theta(\beta)\zeta^{*}],
\label{expectation-value-Q-2}
\end{eqnarray}
\begin{eqnarray}
(Q^{2})
&=&
\left.
-\frac{\partial^{2}}{\partial q^{2}}(\mbox{CF})
\right|_{\gamma=\gamma^{*}=\gamma'=\gamma'^{*}=0} \nonumber \\
&=&
\frac{\hbar}{2\lambda}
[\cosh^{2}\theta(\beta)+\sinh^{2}\theta(\beta) \nonumber \\
&&
+(\cosh\theta(\beta)\alpha^{*}+\sinh\theta(\beta)\zeta
+\cosh\theta(\beta)\alpha+\sinh\theta(\beta)\zeta^{*})^{2}],
\end{eqnarray}
and we obtain:
\begin{equation}
(\Delta Q)^{2}
=
\frac{\hbar}{2\lambda}
[\cosh^{2}\theta(\beta)+\sinh^{2}\theta(\beta)],
\label{Delta-Q-square-0}
\end{equation}
where uncertainty in an arbitrary operator $A$ is given by Eq.~(\ref{definition-operator-uncertainty-0}).

From Eq.~(\ref{characteristic-function-B}),
the expectation values of $P$ and $P^{2}$ are given by:
\begin{eqnarray}
(P)
&=&
\left.
i\frac{\partial}{\partial p}(\mbox{CF})
\right|_{\gamma=\gamma^{*}=\gamma'=\gamma'^{*}=0} \nonumber \\
&=&
i\sqrt{\frac{\lambda\hbar}{2}}
[\cosh\theta(\beta)\alpha^{*}+\sinh\theta(\beta)\zeta
-\cosh\theta(\beta)\alpha-\sinh\theta(\beta)\zeta^{*}],
\label{expectation-value-P-2}
\end{eqnarray}
\begin{eqnarray}
(P^{2})
&=&
\left.
-\frac{\partial^{2}}{\partial q^{2}}(\mbox{CF})
\right|_{\gamma=\gamma^{*}=\gamma'=\gamma'^{*}=0} \nonumber \\
&=&
\frac{\lambda\hbar}{2}
[\cosh^{2}\theta(\beta)+\sinh^{2}\theta(\beta) \nonumber \\
&&
-
(\cosh\theta(\beta)\alpha^{*}+\sinh\theta(\beta)\zeta
-\cosh\theta(\beta)\alpha-\sinh\theta(\beta)\zeta^{*})^{2}],
\end{eqnarray}
and we obtain:
\begin{equation}
(\Delta P)^{2}
=
\frac{\lambda\hbar}{2}
[\cosh^{2}\theta(\beta)+\sinh^{2}\theta(\beta)].
\label{Delta-P-square-0}
\end{equation}

Putting the above results together,
we have Eq.~(\ref{uncertainity-relation-0}) as the uncertainty relation.
The uncertainty relation of Eq.~(\ref{uncertainity-relation-0})
holds not only for $|\alpha,\zeta;\beta)$ but also for $|\alpha,\zeta;\beta\rangle$ and $||\alpha,\zeta;\beta\rangle\!\rangle$.
This is because $|\alpha,\zeta;\beta\rangle$ can be obtained by replacing $\alpha$ and $\zeta$ of $|\alpha,\zeta;\beta)$
according to Eq.~(\ref{replacement-formula-thermal-coherent-states-B}),
$||\alpha,\zeta;\beta\rangle\!\rangle$ can be obtained by replacing $\alpha$ and $\zeta$
of $|\alpha,\zeta;\beta)$
according to Eq.~(\ref{alpha-replacement-1}),
and Eq.~(\ref{uncertainity-relation-0}) does not rely on $\alpha$ or $\zeta$.

At the end of this section,
we write down the expectation values of $Q$ and $P$ for $||\alpha,\zeta;\beta\rangle\!\rangle$.
Replacing $\alpha$ and $\zeta$
of expectation values $(Q)$ and $(P)$
given by Eqs.~(\ref{expectation-value-Q-2}) and (\ref{expectation-value-P-2})
according to Eq.~(\ref{alpha-replacement-1}),
we obtain:
\begin{equation}
\langle\!\langle Q\rangle\!\rangle
=
\sqrt{\frac{\hbar}{2\lambda}}
(\alpha+\alpha^{*}),
\label{expectation-value-Q-3}
\end{equation}
\begin{equation}
\langle\!\langle P\rangle\!\rangle
=
i\sqrt{\frac{\lambda\hbar}{2}}
(\alpha^{*}-\alpha).
\label{expectation-value-P-3}
\end{equation}

\section{\label{section-Appendix-B}Quasiprobability distributions
of $|\alpha,\zeta;\beta)$ and
\\
$||\alpha,\zeta;\beta\rangle\!\rangle$}
\subsection{\label{section-P-representation-conventional}The $P$-representations of $|\alpha,\zeta;\beta)$
and $||\alpha,\zeta;\beta\rangle\!\rangle$}
In this subsection,
we derive the $P$-representations
of $|\alpha,\zeta;\beta)$ and $||\alpha,\zeta;\beta\rangle\!\rangle$.
First,
we compute the $P$-representation of $|\alpha,\zeta;\beta)$.
Its definition is given as follows \cite{Walls1994,Barnett1997}:
\begin{eqnarray}
\rho
&=&
\mbox{Tr}_{\tilde{\cal H}}(|\alpha,\zeta;\beta)(\alpha,\zeta;\beta|) \nonumber \\
&=&
\int d^{2}\mu\,
P(\mu)|\mu\rangle\langle\mu|,
\label{definition-P-representation-1}
\end{eqnarray}
\begin{equation}
P(\mu)
=
\frac{1}{\pi^{2}}
\int d^{2}\eta
\exp(\mu\eta^{*}-\mu^{*}\eta+\frac{1}{2}|\eta|^{2})
\chi(\eta),
\label{definition-P-representation-2}
\end{equation}
\begin{equation}
\chi(\eta)
=
\mbox{Tr}\{\rho \exp(\eta a^{\dagger}-\eta^{*}a)\}.
\label{definition-P-representation-3}
\end{equation}
The characteristic function $\chi(\eta)$ can be rewritten as:
\begin{eqnarray}
\chi(\eta)
&=&
(\alpha,\zeta;\beta|e^{\eta a^{\dagger}-\eta^{*}a}|\alpha,\zeta;\beta) \nonumber \\
&=&
\left.
(\mbox{CF})
\right|_{\gamma=\eta,\gamma^{*}=\eta^{*},\gamma'=\gamma'^{*}=0},
\label{characteristic-function-chi-2}
\end{eqnarray}
where $(\mbox{CF})$ is given by Eq.~(\ref{characteristic-function-B}).

Moreover,
applying Eq.~(\ref{tilde-conjugation-2}) to Eq.~(\ref{characteristic-function-chi-2})
for the invariance under the tilde conjugation,
we obtain:
\begin{equation}
\chi(\eta)
=
\exp[
-\frac{1}{2}(\cosh^{2}\theta(\beta)+\sinh^{2}\theta(\beta))|\eta|^{2}
+(\cosh\theta(\beta)+\sinh\theta(\beta))(\alpha^{*}\eta-\alpha\eta^{*})
].
\label{characteristic-function-C}
\end{equation}
Thus,
we can compute $P(\mu)$ as:
\begin{equation}
P(\mu)
=
G(\mu;
[\cosh\theta(\beta)+\sinh\theta(\beta)]\alpha,
\frac{1}{\sqrt{2}}\sinh\theta(\beta)),
\label{P-representation-final-2}
\end{equation}
where $G(x;\bar{x},\sigma)$ is given by Eq.~(\ref{Gaussian-function}).

Next,
we derive $P(\mu)$ of $||\alpha,\alpha^{*};\beta\rangle\!\rangle$.
Performing a replacement,
\begin{equation}
\alpha
\to
\alpha[\cosh\theta(\beta)-\sinh\theta(\beta)],
\label{alpha-replacement}
\end{equation}
to $P(\mu)$ of $|\alpha,\alpha^{*};\beta)$ given by Eq.~(\ref{P-representation-final-2}),
we obtain:
\begin{equation}
P(\mu)
=
G(\mu;
\alpha,
\frac{1}{\sqrt{2}}\sinh\theta(\beta)),
\label{P-representation-final-3}
\end{equation}
where Eq.~(\ref{alpha-replacement}) is obtained by applying Eq.~(\ref{tilde-conjugation-2})
to Eq.~(\ref{alpha-replacement-1}).

Taking the limit
$\theta(\beta)\to 0$
in Eq.~(\ref{P-representation-final-2}),
we obtain:
\begin{eqnarray}
\lim_{\theta(\beta)\to 0}P(\mu)
&=&
\lim_{\theta(\beta)\to 0}
\frac{1}{\pi\theta^{2}(\beta)}
\exp[-\frac{1}{\theta^{2}(\beta)}|\alpha-\mu|^{2}] \nonumber \\
&=&
\delta^{(2)}(\alpha-\mu).
\label{limit-P-representation-1}
\end{eqnarray}
In the derivation of Eq.~(\ref{limit-P-representation-1}),
we utilize the following formula \cite{Barnett1997}:
\begin{equation}
\lim_{\epsilon\to 0}
\frac{1}{\sqrt{\pi}\epsilon}\exp(-\frac{x^{2}}{\epsilon^{2}})
=
\delta(x).
\end{equation}
It is well known that the $P$-representation of the coherent state $|\alpha\rangle$ is given by
$P(\mu)=\delta^{(2)}(\alpha-\mu)$.
From Eqs.~(\ref{P-representation-final-1}) and (\ref{P-representation-final-3}),
we notice that similar relations to Eq.~(\ref{limit-P-representation-1})
hold for $P(\mu)$ of $||\alpha,\alpha^{*};\beta\rangle\!\rangle$ and $|\alpha,\alpha^{*};\beta\rangle$, as well.

At the end of this subsection,
we discuss the completeness of $|\alpha,\alpha^{*};\beta)$, $||\alpha,\alpha^{*};\beta\rangle\!\rangle$ and $|\alpha,\alpha^{*};\beta\rangle$.
First,
we show that $|\alpha,\alpha^{*};\beta)$ forms a complete system.
From Eq.~(\ref{P-representation-final-2}),
we obtain:
\begin{equation}
\int d^{2}\alpha\,P(\mu)
=
\frac{1}{[\cosh\theta(\beta)+\sinh\theta(\beta)]^{2}}.
\end{equation}
Thus,
setting
\begin{equation}
\rho(\alpha,\alpha^{*};\beta)
=
\mbox{Tr}_{\tilde{\cal H}}(|\alpha,\alpha^{*};\beta)(\alpha,\alpha^{*};\beta|),
\label{density-operator-H-space-2}
\end{equation}
we obtain:
\begin{equation}
\frac{1}{\pi}[\cosh\theta(\beta)+\sinh\theta(\beta)]^{2}
\int d^{2}\alpha\,\rho(\alpha,\alpha^{*};\beta)
=
\mbox{\boldmath $I$}.
\end{equation}
Hence,
we confirm that $\{\rho(\alpha,\alpha^{*};\beta):\alpha\in\mathbf{C}\}$
given by Eq.~(\ref{density-operator-H-space-2}) forms a complete system.
Similar relations hold for  $||\alpha,\alpha^{*};\beta\rangle\!\rangle$ and $|\alpha,\alpha^{*};\beta\rangle$,
so that we can confirm the completeness of  $||\alpha,\alpha^{*};\beta\rangle\!\rangle$ and $|\alpha,\alpha^{*};\beta\rangle$.

\subsection{\label{section-Q-representation-conventional}The $Q$-functions and the Wigner functions of $|\alpha,\alpha^{*};\beta)$
and $||\alpha,\alpha^{*};\beta\rangle\!\rangle$}
In this subsection,
we derive the $Q$-functions and the Wigner functions of $|\alpha,\alpha^{*};\beta)$ and $||\alpha,\alpha^{*};\beta\rangle\!\rangle$.
First of all,
we compute $Q$-function of $|\alpha,\alpha^{*};\beta)$.
It is defined as follows \cite{Walls1994,Barnett1997}:
\begin{eqnarray}
Q(\mu)
&=&
\frac{1}{\pi}\langle\mu|\rho|\mu\rangle \nonumber \\
&=&
\frac{1}{\pi^{2}}\int d^{2}\eta
\exp(\mu\eta^{*}-\mu^{*}\eta-\frac{1}{2}|\eta|^{2})
\chi(\eta),
\label{Q-representation-definition-1}
\end{eqnarray}
where $\chi(\eta)$ is given by Eq.~(\ref{characteristic-function-C}).
Thus,
we obtain:
\begin{equation}
Q(\mu)
=
G(\mu;
[\cosh\theta(\beta)+\sinh\theta(\beta)]\alpha,
\frac{1}{\sqrt{2}}\cosh\theta(\beta)).
\label{Q-representation-final-2}
\end{equation}

To obtain $Q(\mu)$ of $||\alpha,\alpha^{*};\beta\rangle\!\rangle$,
we only have to perform the replacement in accordance with Eq.~(\ref{alpha-replacement})
to $Q(\mu)$ of $|\alpha,\alpha^{*};\beta)$ given by Eq.~(\ref{Q-representation-final-2}),
and we obtain:
\begin{equation}
Q(\mu)
=
G(\mu;
\alpha,
\frac{1}{\sqrt{2}}\cosh\theta(\beta)).
\label{Q-representation-final-3}
\end{equation}

Next,
we derive the Wigner functions of $|\alpha,\alpha^{*};\beta)$ and $||\alpha,\alpha^{*};\beta\rangle\!\rangle$.
First,
we compute the Wigner function of $|\alpha,\alpha^{*};\beta)$.
Using the characteristic function $\chi(\eta)$ given by Eq.~(\ref{characteristic-function-C}),
we define $W(\mu)$ as follows \cite{Walls1994,Barnett1997}:
\begin{equation}
W(\mu)
=
\frac{1}{\pi^{2}}
\int d^{2}\eta\,
\chi(\eta)\exp(-\eta\mu^{*}+\eta^{*}\mu).
\label{definition-Wigner-function}
\end{equation}
Thus,
we obtain:
\begin{equation}
W(\mu)
=
G(\mu;
[\cosh\theta(\beta)+\sinh\theta(\beta)]\alpha,
\frac{1}{2}[\cosh^{2}\theta(\beta)+\sinh^{2}\theta(\beta)]^{1/2}).
\label{W-representation-final-2}
\end{equation}

To obtain $W(\mu)$ of $||\alpha,\alpha^{*};\beta\rangle\!\rangle$,
we only have to perform the replacement in accordance with
Eq.~(\ref{alpha-replacement})
to $W(\mu)$ of $|\alpha,\alpha^{*};\beta)$ given by Eq.~(\ref{W-representation-final-2}),
and we obtain:
\begin{equation}
W(\mu)
=
G(\mu;
\alpha,
\frac{1}{2}[\cosh^{2}\theta(\beta)+\sinh^{2}\theta(\beta)]^{1/2}).
\label{W-representation-final-3}
\end{equation}

\end{document}